\documentclass[fleqn,usenatbib]{mnras}

\usepackage{bm}

\usepackage[T1]{fontenc}
\usepackage[utf8]{inputenc}

\DeclareRobustCommand{\VAN}[3]{#2}
\let\VANthebibliography\thebibliography
\def\thebibliography{\DeclareRobustCommand{\VAN}[3]{##3}\VANthebibliography}

\usepackage{graphicx}	
\usepackage{amsmath}	
\usepackage{amssymb}	
\usepackage{ulem}

\newcommand{\cs}{c_{\rm s}}

\newcommand{\LJ}{L_{\rm J}}
\newcommand{\LJp}{L_{{\rm J},p}}
\newcommand{\M}{{\cal M}}

\newcommand{\MJ}{M_{\rm J}}

\newcommand{\rhoavg}{\bar\rho}
\newcommand{\s}{{\bm S}}

\newcommand{\T}{{\bm T}}
\newcommand{\uu}{{\bm u}}
\newcommand{\xx}{{\bm x}}

\newcommand{\dif}{\mathrm{d}}
\newcommand{\roverr}{ \left( \frac{r}{r_0} \right) }

\title[Density evolution during core collapse]{Density profile evolution during prestellar core collapse:
Collapse starts at the large scale}

\author[Gómez, Vázquez-Semadeni, \& Palau]{
Gilberto C. Gómez,\thanks{E-mail: g.gomez@irya.unam.mx},
Enrique Vázquez-Semadeni,
 and
Aina Palau
\\
$^{1}$Instituto de Radioastronomía y Astrofísica,
      Universidad Nacional Aut\'onoma de M\'exico, Apdo. postal 3-72,
      Morelia Mich. 58089, M\'exico
}

\date{Accepted XXX. Received YYY; in original form ZZZ}

\pubyear{2020}

\begin{document}
\label{firstpage}
\pagerange{\pageref{firstpage}--\pageref{lastpage}}
\maketitle

\begin{abstract}

We study the gravitationally-dominated, accretion-driven evolution of a prestellar core. In our model, as the core's density increases, it remains immersed in a constant-density environment and so it accretes from this environment, increasing its mass and reducing its Jeans length. Assuming a power-law density profile $\rho \propto r^{-p}$, we compute the rate of change of the slope $p$, and show that the value $p=2$ is stationary, and furthermore, an attractor. The radial profile of the Jeans length scales as $r^{p/2}$, implying that, for $p<2$, there is a radius below which the region is smaller than its Jeans length, thus appearing gravitationally stable and in need of pressure confinement, while, in reality, it is part of a larger-scale collapse and is undergoing compression by the infalling material. In this region, the infall speed decreases towards the center, eventually becoming subsonic, thus appearing ``coherent'', without the need for turbulence dissipation. We present a compilation of observational determinations of density profiles in dense cores and show that the distribution of their slopes peaks at $p \sim 1.7$--1.9, supporting the notion that the profile steepens over time. Finally, we discuss the case of magnetic support in a core in which the field scales as $B \propto \rho^\beta$. For the expected value of $\beta = 2/3$, this implies that the mass to magnetic flux ratio also decreases towards the central parts of the cores, making them appear magnetically supported, while in reality they may be part of larger collapsing supercritical region. We conclude that local signatures of either thermal or magnetic support are not conclusive evidence of stability, that the gravitational instability of a region must be established at the large scales, and that the prestellar stage of collapse is dynamic rather than quasistatic.


\end{abstract}




\section{Introduction} \label{sec:intro}

When studying dense objects, ranging from molecular clouds to dense cores, it is standard practice to measure their masses, sizes, and temperatures (or velocity dispersions, in general), to determine whether they are gravitationally bound or unbound. When they appear unbound, it is therefore customary to assume that they are confined by some external pressure \citep[e.g.,] [] {Keto_Myers86, Lada+08, Field+11, Leroy+15, Kirk+17, Chen+19}. The external pressure is often interpreted as being  caused by the weight of the surrounding material, but still of {\it confining} (i.e., hydrostatic) nature. 

However, this line of reasoning is somewhat flawed in the sense that the objects being considered are {\it already} dense. That is, they must have arrived at their high density by some mechanism. In the {\it inside-out} collapse model of \citet{Shu77}, the prestellar stage of collapse should occur quasistatically, perhaps supported by magnetic fields and the contraction occurring through ambipolar diffusion \citep[] [see further discussion below] {Mestel_Spitzer56}. In the more recent and widespread {\it gravoturbulent} paradigm \citep[e.g.,] [] {VS+03, MacLow_Klessen04, BP+07} it is assumed that the dense cores within clouds reach their densities by means of supersonic turbulent compressions, and that they subsequently may or may not undergo collapse depending on whether their density becomes large enough to become locally gravitationally unstable \citep{Galvan+07}; i.e., their Jeans mass (see eq.\ [\ref{eq:MJ}] below) becomes smaller than their own physical mass due to the compression. At the same time, the gravoturbulent assumption is that the parent structures (the clouds) are supported against collapse by the turbulent pressure. However, another possibility is that the cores have their already large densities as a result of already ongoing gravitational contraction which started earlier in a larger-scale and lower-density structure. In this paper we argue in favor of this possibility.

Gravitational contraction constitutes the fundamental mechanism of structure formation in the Universe, and the instability analysis by \citet{Jeans1902} is the starting building block for the onset of gravitational contraction in a self-gravitating medium with thermal pressure support. Instability in a uniform isothermal medium with density $\rho$ and sound speed $\cs$ occurs for perturbations of wavelength larger than the {\it Jeans length}, given by
\begin{equation}
    \LJ \equiv \left(\frac{\pi \cs^2} {G \rho} \right)^{1/2}.
    \label{eq:LJ}
\end{equation}
From this size scale, it is customary to define the {\it Jeans mass} as the mass of a spherical gas cloud of uniform density whose radius equals half the Jeans length, so that
\begin{equation}
    \MJ \equiv \frac{4}{3}\pi \rho \left(\frac{\LJ}{2}\right)^3 = \frac{\pi^{5/2} \cs^3} {6 G^{3/2} \rho^{1/2}},
    \label{eq:MJ}
\end{equation}
so that clouds of mass $M>\MJ$ are unstable to gravitational collapse.

The subsequent evolution of the collapsing structure has been extensively investigated both analytically and numerically. Already by the mid XX-th century, \citet{Hoyle53} made the very important point that, for a cloud subject to cooling in such a way that it remains roughly isothermal (or, in general, such that its pressure scales as $P \propto \rho^\gamma$, with $\gamma < 4/3$), the Jeans mass {\it decreases} during the contraction, allowing for the possibility of {\it gravitational fragmentation} of the collapsing mass. He also found that the fragmentation continues until the cloud becomes optically thick, so that it traps the heat released by the collapse and its thermal behavior becomes closer to adiabatic, a result that continues to be confirmed today regardless of other properties of the cloud, such as its turbulent state \citep[e.g.,] [] {Guszejnov+18, Lee_Henneb18}.

The density and velocity profiles of a collapsing spherical mass were studied intensely using similarity methods \citep[e.g.,] [] {Larson69, Penston69, Shu77, Hunter77, Whitworth+85}. The latter authors, hereafter WS85, provided a compendium of the various possible similarity collapse regimes depending on the parameters of the problem. However, similarity solutions are precluded by nature from modeling the initial transients that lead from the ad hoc initial conditions to the similarity solution. Numerical simulations are in general needed for this task.

The early numerical studies of \citet{Larson69} and \citet{Penston69} (hereafter referred to as the LP solution) found that the prestellar stage settled to a solution characterized by an inner region with a flat density profile and an infall velocity profile linear with radius, and an outer region with an $r^{-2}$ density profile and a uniform, supersonic infall velocity. This configuration is consistent with the asymptotic forms of the similarity equations they derived. WS85 also found this asymptotic solution, as one of multiple possible solutions, depending on the system's parameters.

Perhaps the most famous similarity solution for the collapse problem is the so-called {\it inside-out} collapse proposed by \citet[] [hereafter Shu77] {Shu77}. This solution corresponds to the {\it protostellar} stage of collapse---i.e., after the singularity (the star) has formed---, since its ``initial condition'' is the hydrostatic solution of a singular isothermal sphere (SIS),\footnote{Strictly speaking, a similarity model cannot represent the transition from a fully hydrostatic state to a dynamical one. The SIS is considered an "initial" condition for the inside-out collapse solution in the sense that the inside-out solution of Shu77 has zero velocity at large values of the similarity variable $x=r/\cs t$, and so, it has zero velocity {\it almost} everywhere as $t\rightarrow 0$. Nevertheless, the actual transition from zero velocity strictly everywhere to the inside-out solution is outside the realm of a similarity study.} characterized by a density profile $\rho \propto r^{-2}$ and zero infall speed, $v=0$, everywhere. Since the density diverges at the center of an SIS, this ``initial condition'' corresponds to the time at which a protostellar object appears. In this solution, the core has density and infall velocity profiles given by $\rho \propto r^{-3/2}$ and $-v \propto r^{-1/2}$, respectively, out to a rarefaction front. Beyond this, the profile is like that of the SIS, with $\rho \propto r^{-2}$ and $v =0$.

Shu77 argued that the {\it prestellar} stage (i.e., the evolution {\it before} the protostar forms, or the approach to the SIS) should occur quasistatically, rather than dynamically. He reasoned that, in order to reach such a configuration, detailed mechanical balance between the thermal pressure gradient and self-gravity would be necessary, in analogy with the general Bonnor-Ebert (BE)-sphere \citep{Ebert55, Bonnor56} hydrostatic solution. He also suggested that the LP solution was unrealistic, arguing that only finely-tuned initial and boundary conditions could lead to it. However, numerical simulations of self-consistently evolving cores from non-singular initial conditions systematically show that the flow approaches the LP solution \citep[e.g.,] [although see \citealt{Keto+15} for a counterexample] {Larson69, Penston69, Hunter77, Foster_Cheval93, Moham_Stahler13, NaranjoRomero+15}. In addition, Shu's inside-out solution has a number of problems of its own. First, its initial condition, the SIS, constitutes the most unstable possible hydrostatic solution for a spherical core, and is therefore extremely unlikely, or plain impossible,
to self-consistently develop in turbulent molecular clouds \citep{Whitworth+96}. Second, if a quasistatic configuration were to develop, for example, by slow contraction mediated by ambipolar diffusion, the resulting object would be highly flattened rather than spherical, and would have a finite, rather than singular, central density. Third, observations are generally inconsistent with the inside-out collapse, exhibiting infall motions that extend beyond the expected location of the rarefaction front for inside-out collapse \citep[e.g.,] [] {Lee+01} and line profiles that do not match those obtained from the SIS \citep[e.g.,][]{Keto+15, Koumpia+20}.

An important contribution in this regard was made by \citet{Li18}, who showed that an $r^{-2}$ density profile follows simply from the assumption of spherical free-fall collapse, under the conditions that the infall speed at every radius is just the gravitational speed $\sqrt{\eta GM/r}$, where $\eta$ is a geometrical constant, and that the mass flux across spherical shells is constant, independent of radius. This implies that the $r^{-2}$ density profile does not require detailed mechanical balance nor quasistatic contraction, and can originate simply from unimpeded gravitationally-driven flow.

In this paper we now investigate the transient approach to an $r^{-2}$ density profile, in particular in a collapsing region within a uniform medium, inspired by the results of \citet[] [hereafter Paper I] {NaranjoRomero+15}, who modeled the growth of a Jeans-mass density fluctuation embedded in a uniform medium. The setup in Paper I attempted to represent the mechanism of global hierarchical collapse \citep[GHC;] [] {VS+19}, in which local collapses begin to occur as a consequence of the large-scale gravitational contraction of the parent cloud, causing a reduction of the Jeans mass. Therefore, fluctuations of a certain mass $M$ become unstable when the mean Jeans mass in the cloud becomes smaller than $M$. Thus, the setup in Paper I represented the onset of gravitational collapse of a fluctuation of mass $M$ when the mean Jeans mass in the contracting parent cloud becomes smaller than $M$. Note that in this scenario, the large-scale contraction is directed to a distant collapse center, different from the local collapse center of the fluctuation, in a {\it conveyor-belt} fashion \citep{Longmore+14}. Therefore, this mode of collapse is modeled by the local collapse of structure of roughly one Jeans mass within a globally Jeans-unstable substrate \citep{VS+19}.

In the simulation of Paper I, it was observed that the density profile steepened as time progressed, approaching the $r^{-2}$ profile characteristic of the LP solution at large distances from the center, while remaining flat at the center. Here we use a simplified analytical description simply assuming that the fluctuation evolves along a series of power laws of the form $\rho \propto r^{-p}$, with $0<p<3$. This simplification neglects the central flat-density part of the core, but we consider it introduces no significant error, since the mass interior to a radius $r$ vanishes as $r \rightarrow 0$ for $p<3$. We then show explicitly that the Jeans length decreases with time as the profile steepens, and that the slope $2$ is actually an {\it attractor}; i.e., values different from $2$ imply a change of $p$ that approaches that value.

The plan of the paper is as follows. In Sec. \ref{sec:Jeans} we introduce the model, and compute the Jeans length as a function of $p$, and show that it shrinks to zero at $p=2$. In Sec.\ \ref{sec:p_evol} we then compute the evolution of $p$, showing that it approaches 2. Next, in Sec.\ \ref{sec:discussion} we discuss some implications of our results, in particular concerning the nature of cores that locally appear gravitationally unbound. Finally, in Sec.\ \ref{sec:concls} we give a summary and some conclusions.

\section{The contraction of the Jeans length} \label{sec:Jeans}

\begin{figure}
    \centering
    \includegraphics[width = 0.45 \textwidth]{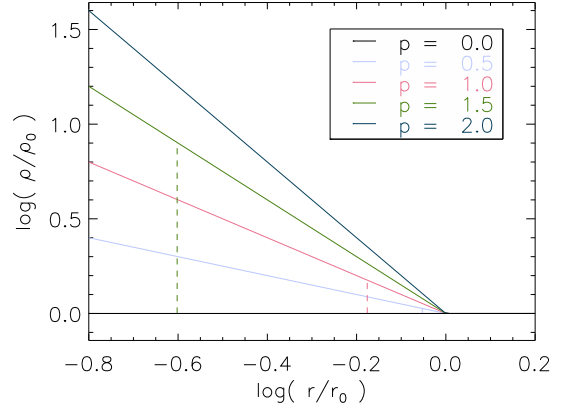}
    \caption{
    Density profile model ({\it solid lines}) and position at which
    one Jeans length is reached ({\it dashed lines}) for a range logarithmic slopes $p$.
    Since $r_0$ and $\rho_0$ are set by the initial Jeans length and surrounding density,
    respectively, the mass within $r_0$ increases as $p$ increases from $0$.
    So, the contracting region accretes from its environment.}
    \label{fig:profile}
\end{figure}

Let us consider a spherically symmetric density distribution (a core) of the form

\begin{equation}
    \rho = \rho_0 \roverr^{-p},
    \label{eq:dens_prof}
\end{equation}

\noindent
where $\rho_0$ is the region's initial density (uniform, with $p=0$),
and $r_0$ is the Jeans length at $\rho_0$. The density beyond $r_0$ remains uniform at $\rho_0$ throughout the evolution. 
We will consider that the region within $r_0$ is undergoing gravitational collapse,
so that $p$ increases in time.
%
As the collapse proceeds, gas at $r>r_0$ will flow inwards, increasing the mass and the mean density inside
$r_0$,
that is, the core accretes from its surroundings. To model this accretion flow, we fix $r_0$ and $\rho_0$ for all $p$, so that the collapsing region's density profile is anchored at these values (see fig. \ref{fig:profile}), as observed in the numerical simulation of \citet{NaranjoRomero+15}. 

At a given $p$, the mean density internal to a radius $r$ is,
\begin{equation}
    \bar{\rho}_p(r) = \frac{3 \rho_0}{3-p} \roverr^{-p}.
    \label{eq:mean_rho}
\end{equation}
The corresponding Jeans length is,
\begin{equation}
    \LJp(r) = r_0 \left[ \frac{3-p}{3} \roverr^{p} \right]^{1/2}.
    \label{eq:LJp_vs_r}
\end{equation}
It is important to note that, according to eq.\ (\ref{eq:LJp_vs_r}), $\LJp(r) \propto r^{p/2}$. This implies
that there is an inner region within the core for which $r < \LJp(r)$, while outside this region, $r > \LJp(r)$.
Indeed, let us now define $x$ as the number of Jeans lengths within a radius $r$,
\begin{equation}
    x \equiv \frac{r}{\LJp(r)} =\left[ \frac{3}{3-p} \roverr^{2-p} \right]^{1/2}.
\end{equation}
We thus see that, for $0 < p < 2$, as $r$ increases, the number of Jeans lengths contained in the region increases.
Conversely, at fixed $r$, the number of Jeans lengths contained within $r$ increases as $p$ increases from 0
to 2. The radius at which one Jeans length is reached ($x=1$) is,
\begin{equation}
    \frac{r_{x=1}}{r_0} = \left( \frac{3-p}{3} \right)^{1/(2-p)}.
\end{equation}
Note that, as the region evolves and $p$ increases from $0$, $r_{x=1}$
shrinks to zero as $p \rightarrow 2$, in a manner consistent with the fact that such profile will contain a
constant number of Jeans lengths at every radius (exactly one Jeans length at each radius for the specific case
of the SIS). Therefore, the largest scales of the region become
unstable before the smaller ones, and {\it the prestellar collapse proceeds from the outside-in}. 

Note also that, during the whole prestellar transient stage, for which $p<2$, the region $x<1$, in which $r < \LJp(r)$, appears Jeans-stable, in spite of being the centermost part of a larger-scale collapse. 
We discuss this further in Sec.\ \ref{sec:LJ_decrease}.

\section{Collapsing profile evolution} \label{sec:p_evol}

We now explore the way the density profile evolves during the collapse.
Consider the continuity equation,
\begin{equation}
    \frac{\partial \rho}{\partial t} = - \nabla \cdot \left(\rho \mathbf{v}\right).
    \label{eq:continuity}
\end{equation}
Following \citet{Li18}, let us assume that the radial flow velocity at $r<r_0$ is given by
\begin{equation}
    v_r = -f v_\mathrm{ff}, 
    \label{eq:vr}
\end{equation}
where $v_\mathrm{ff}=\sqrt{G M(r)/r}$, 
$M(r)$ is the mass internal to $r$, and $f$ is a constant. Thus,
\begin{equation}
    v_r = -\left[ \frac{4 \pi G \rho_0 r_0^2 f^2}{3-p} \roverr^{2-p} \right]^{1/2}
    \label{eq:v_inflow}
\end{equation}
%
So, the right-hand side of eq.\ (\ref{eq:continuity}) is,
\begin{eqnarray}
    - \nabla \cdot \left(\rho \mathbf{v}\right) &=& -\frac{1}{r^2} \frac{\partial r^2 \rho v_r}{\partial r} \nonumber \\
    &=& \left( 3-\frac{3p}{2}\right) \left[ \frac{4\pi G \rho_0^3 f^2}{(3-p)r_0^2}\right]^{1/2} \roverr^{-3p/2}.
\end{eqnarray}
On the other hand, assuming that the density distribution approximately evolves from one power-law
to another, the left hand side of eq. (\ref{eq:continuity}) is
\begin{equation}
    \frac{\partial \rho}{\partial t} = -\rho_0 \left(\ln\frac{r}{r_0}\right) \roverr^{-p}
                                       \frac{\dif p}{\dif t}.
    \label{eq:drhodt}
\end{equation}
Combining these two equations yields,
\begin{equation}
    \frac{\dif p}{\dif t} = \left( 3-\frac{3p}{2}\right)\left[ \frac{4\pi G \rho_0 f^2}{3-p}\right]^{1/2}
      \frac{(r/r_0)^{-p/2}}{-\ln(r/r_0)}.
  \label{eq:dpdt}
\end{equation}

Since we are considering the region where $r<r_0$,
the right-hand-side sign is given by the $(3-3p/2)$ factor.
Therefore, if $p < 2$, $p$ increases at all radii, while it decreases if $p>2$:
a shallow profile steepens, while a steeper profile flattens, reaching
a steady state when $p=2$, in agreement with the result by \citet{Li18} that the mass flux across spherical shells
is independent of radius at this slope. { Therefore, the value $p=2$ is an ``attractor" for the slope,
under our assumptions that the density profile is fixed at $\rho_0$ at $r_0$ and that the flow is given by the free fall velocity.} This is also in agreement with the conclusion from \citet{Murray_Chang15} that the density profile in protostellar systems (i.e., after a protostar has formed) approaches a time-stationary form, so that $\rho(r,t) \rightarrow \rho(r)$.

%
%
%
%

The region within one Jeans length ($x<1$) may be considered as a prestellar core, which grows in mass as it
accretes from the outer regions. This accretion rate is set by the inflow velocity (eq. \ref{eq:v_inflow}) at $x=1$:
\begin{equation}
    \dot M_p(r_{x=1}) 
      = 3f \left(\frac{G M_0^3}{r_0^3}\right)^{1/2} \left(\frac{3-p}{3}\right)^{1/2},
      \label{eq:Mdotx1}
\end{equation}
where $M_0 = (4\pi/3) r_0^3 \rho_0$ is the Jeans mass at the initial density.
Substituting the Jeans length for $r_0$, eq.\ (\ref{eq:Mdotx1}) reads,
\begin{equation}
    \dot M_p(r_{x=1}) = f \frac{\sqrt{3-p}}{3} \, \frac{\pi^3 c_s^3}{G},
\end{equation}
which is about an order of magnitude larger than the expansion-wave collapse solution
described in \citet{Shu77}. 

\section{Discussion}
\label{sec:discussion}

\subsection{The density profile} \label{sec:rho?prof}

\subsubsection{The temporal decrease of the Jeans length and the inner ram-pressure-compressed region}
\label{sec:LJ_decrease}


Although in this paper we have idealized the evolution of the density profile as a single, evolving power law, we nevertheless recover the fact that, for all $p < 2$ and during the entire prestellar collapse stage, there is always a central region which is smaller than the Jeans length that corresponds to this region's mean density \citep{Whitworth+85}. In the full similarity solution, this central region corresponds to the central flat part of the density profile \citep{Keto+10}. In our approximate evolutionary solution, the density profile continues with the same logarithmic slope all the way to the center, but there is still a region smaller than the Jeans length. Also, as indicated both by our solution as well as by the similarity solution, the mean density of this region increases with time and its physical size decreases, shrinking to zero at the time of the formation of the singularity (the protostar).

The fact that the Jeans length decreases during the collapse of an isothermal region has been known for decades \citep{Hoyle53}, but it is important to recall it, because some of its consequences are often overlooked. For example, when an object which is only marginally unstable (i.e., with a mass $M = (1+\epsilon) \MJ$, where $0<\epsilon \ll 1$) begins to collapse, almost any subregion interior to it is Jeans ``stable'' \citep[see also] [] {Gomez+07, Gong_Ostr09}. Yet, the entire object is collapsing. As the collapse advances, although the Jeans length decreases, any region interior to it continues to appear stable, even though the density of this region is increasing and its size is decreasing. That is, this Jeans-stable central region is contracting because it is being crushed by the infall of the large-scale, Jeans unstable whole. However, if this central object is observed in isolation, it can be confused with being in equilibrium. This may well be the case of apparently pressure-confined cores such as those of \citet{Lada+08}, \citet{Kirk+17}, and \citet{Chen+19}. 

Indeed, it is noteworthy that \citet{Lada+08} found that the apparently stable pressure-confined, BE-like cores in the Pipe cloud fall in the same locus as the unstable ones in the diagram of $M_{\rm core}/M_{\rm BE}$ {\it vs.} $M_{\rm core}$, where $M_{\rm core}$ is the mass of the core and $M_{\rm BE}$ is the Bonnor-Ebert mass corresponding to the core's mean density and temperature. At face value, this would be a surprising result, as there is no obvious reason why stable, hydrostatic, pressure-confined cores should occupy the same locus as the unstable, dynamically collapsing ones. An explanation was provided by \citet{NaranjoRomero+15}, who showed that, in their simulations of the collapse of a spherical Gaussian perturbation on top of a uniform, globally-unstable medium, regions defined by a certain density threshold above the background density appeared Jeans-stable at early stages of the collapse, yet occupied the same locus as the later, clearly unstable stages. They thus suggested that the apparent stability was just due to the failure to recognize that the core was just the innermost part of a globally unstable larger-scale object, being compressed by it.

Similarly, \citet{Kirk+17} found that most of the dense ammonia cores in Orion appear to be gravitationally unbound when considering only their self-gravity and internal pressure, but that they appear bound when the external pressure is considered. \citet{Chen+19} found similar results for the L1688 region of Ophiucus and the B18 region of Taurus. Moreover, for the external pressure, \citet{Kirk+17} included the contribution of the nonthermal velocity dispersion. Although the latter is usually interpreted as turbulence, an equally valid alternative interpretation is that it corresponds to the ram pressure produced by the infall of the surrounding envelope, as indicated by our calculations and the simulation of \citet{NaranjoRomero+15}. In this case, the cores are not just pressure confined, but rather they are being ram-pressure-compressed.


It is noteworthy that, since the ram-pressure confinement applies throughout the clump, because gas is continuously accreting, it is to be expected that the critical mass for stability  be lower than the traditional Bonnor-Ebert mass, as suggested by \citet{Hunter_Fleck82}.

\subsubsection{The density profile and geometry} \label{sec:geometry}

The origin of the slope of the density profile is an extremely important consideration. As mentioned in Sec.\ \ref{sec:intro}, \citet{Shu77} suggested that the $r^{-2}$ density profile would be reached ``\dots as long as the initial conditions allow the early phases of the flow to occur subsonically'' since, he argued, this profile is the result of detailed mechanical balance at all radii in the core. However, in this contribution we have shown that it occurs spontaneously during non-homologous spherical gravitational contraction as a consequence of the velocity at every radius being driven by the gravitational attraction of the material internal to it, under the constraint that the mass flux across spherical shells is independent of radius \citep{Li18}, and that other slopes cause a radial mass flux gradient that tends to cancel the gradient. This can be seen from the $3-3p/2$ factor in equation (\ref{eq:dpdt}), which defines $p=2$ as a ``special'' slope, with different slopes evolving towards it. This $p$ value originates from the
geometrical focusing involved in the $r^2$ factors in the divergence operator.\
\footnote{ As an example of the effect of geometry on the attractor slope, for a
cylindrically-symmetric density distribution, i.e. a filament, the infall velocity given by $v = -f  v_\mathrm{ff}=-f \sqrt{G \lambda(R)}$, where $\lambda(R) = \int_0^R 2\pi R \rho \dif R$ is the linear mass density, shows the same radial dependency as in the spherical case, $v \propto R^{(2-p)/2}$. But, for this cylindrical distribution, the equation equivalent to eq.\ (\ref{eq:dpdt}) reads
\begin{equation*}
    \frac{\dif p}{\dif t} = \left( 2-\frac{3p}{2}\right)\left[ \frac{2\pi G \rho_0 f^2}{2-p}\right]^{1/2}
      \frac{(R/R_0)^{-p/2}}{-\ln(R/R_0)},
  \label{eq:dpdt_cyl}
\end{equation*}
%
and the equilibrium slope is $p=4/3$, with steepening or shallowing profiles for smaller or larger $p$ values.

\citet{Li18} obtained the $p=2$ slope for spherical geometry by requiring uniform radial accretion. For cylindrical accretion, such requirement also yields a logarithmic slope of $p=4/3$.}

\subsubsection{Implications of the $r^{-2}$ density profile and its evolution} \label{sec:implications}

It is well known that an $r^{-2}$ density distribution of the form of eq.~(\ref{eq:dens_prof}) with $p=2$ implies that the gravitational potential is given by
\begin{equation}
\varphi(r) = 4 \pi G \rho_0 r_0^2 \ln{\left(\frac{r} {r_0} \right)},
\label{eq:grav_pot}
\end{equation}
and the force is given by
\begin{equation}
    F(r) = -\frac{4 \pi G \rho_0 r_0^3} {r}.
    \label{eq:force}
\end{equation}
That is, for an $r^{-2}$ density profile, both the gravitational potential and the force vary much more slowly with distance than for a point mass. This reinforces the notion that the environment of cores is most likely gravitationally bound to them, a possibility mentioned by \citet{Kirk+17} in regards to their sample of cores in the Orion A cloud.

\subsubsection{Comparison to observations} \label{sec:observations}

In Sec.\ \ref{sec:Jeans} we showed that, for any density profile with $0 < p < 2$, there is always a region at the center of the collapse whose size is smaller than the Jeans length for the mean density within that region, so that it appears Jeans-stable.
As a matter of fact, for an initial density of 10$^3$~cm$^{-3}$ at a temperature of 20~K, the corresponding Jeans length or $r_0$ is $\sim1$~pc. 
Adopting a typical value for $p$ of 1.8 (see below), for such a core the radius at which one Jeans length is reached, $r_{x=1}$, would be about 2000~au. This size is consistent with the size of the compact sources (necessarily resulting from a collapse process) detected by \citet{Huelamo+17}, which are embedded in apparently stable cores.

\begin{figure}
    \centering
    \includegraphics[width = 0.45 \textwidth]{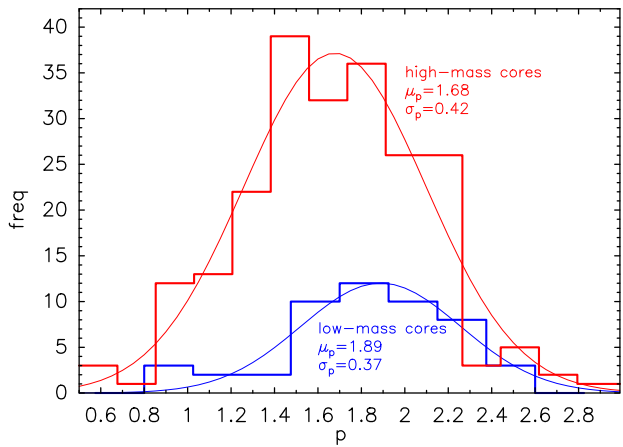}
    \caption{
    Histogram of the measured density power-law indices of cores undergoing star formation in a number of works in the literature (see main text for references). The blue histogram corresponds to low-mass cores and the red histogram corresponds to high-mass cores. The thin curves correspond to Gaussian fits, for which we report the position of the peak, $\mu$, and the standard deviation, $\sigma$.}
    \label{fig:histo}
\end{figure}

Moreover, observational determinations of density profiles in { samples of} dense cores often suggest slopes $p \lesssim 2$. 
In Fig.~\ref{fig:histo} we present a histogram of the $p$ values for a compilation of  different samples of low-mass \citep{ChandlerRicher00, HogerheijdeSandell00, Shirley+00, Shirley+02, MotteAndre01, Young+03} and high-mass cores\footnote{The work of \citet{Williams+05} is not included here because they only test discrete values of $p$: 0.0, 0.5, 1.0, 1.5, 2.0 and 2.5. The work of \citet{Friesen+18} is neither included because these authors concentrate on much smaller scales ($\sim 0.001$~pc).} \citep{vanderTak+00, Beuther+02, Mueller+02, HatchellvanderTak03, Pirogov09, ButlerTan12, Palau+14, Palau+20, Wyrowski+16, Li+19}.
In these works, the typical sampled scales are 0.02--0.2 pc for the low-mass cores and 0.1--1 pc for the high-mass cores. { For the high-mass case we have additionally included the results from Gieser et al. (2020, submitted) studying cores at $\sim0.02$~pc scales.} As can be seen in the figure, the center of the fitted Gaussian for both the low-mass and the high-mass cores is smaller than 2 (1.89 for low-mass cores and { 1.68} for the high-mass cores), fully consistent with the theoretical work presented here.

In our compilation of density power-law indices, a fraction of $\sim6$\% of the cores have $p$ values $>2.3$. We consider here that those with $2 \la p < 2.3$ are consistent with $p=2$, as typical uncertainties in $p$ are around 0.3 \citep[e.g.,][]{Shirley+00}.
Close inspection of the corresponding uncertainties for some of the cores with $p>2.3$ shows that they are large, $>0.6$ \citep[e.g.,][]{MotteAndre01}. Also, in the sample of \citet{Palau+14}, the core with the steepest density power-law index, of 2.45, was reported to have a more accurate value, with $p<2$, in the follow-up work of \citet{Palau+15}. { A few exceptions are reported in the literature of cores with density power law indices as steep as 2.7, which seem to be robustly measured \citep{Didelon+15}, but these are attributed to particular conditions of external compression and/or departure from spherical symmetry \citep{TrevinoMorales+19}. Thus, to our knowledge there is no clear observational evidence of a significant  number of cores presenting density power-law indices steeper than 2.}

The samples used to build the histograms presented in Fig.~\ref{fig:histo} include essentially cores already undergoing star formation. The fact that protostellar cores typically exhibit slopes smaller than 2 may seem conflicting with the result from numerical simulations suggesting that a slope of 2 is reached precisely at the time of formation of the singularity \citep{NaranjoRomero+15}. However, this apparent inconsistency may probably be resolved by the fact that the geometry of actual collapsing core is much more complex than the spherical symmetry assumed here and in those numerical simulations. This may also  ocassionally cause profiles that may appear steeper than 2 when viewed from some particular direction.

For pre-stellar cores, a large number of works have reported shallower density structures, with $p \sim 1$ or following Bonnor-Ebert spheres \citep[e.g.,][]{Evans+01, Shirley+05, Schnee+10, Chen+19}. 
For these shallower profiles, the variation of the potential and the force with distance is even slower. The long range of the gravitational force and potential in this case suggests that it is important to investigate the boundedness of the environment of these structures.

Therefore, the density structure of pre-stellar cores seems to be flatter than the density structure of star-forming cores. This steepening of the density profile as a core evolves through its collapse has been explicitly reported in a number of cases \citep[e.g.,][]{ChandlerRicher00, Beuther+02, Williams+05, Hung+10, Giannetti+13, Gerner+15, Guzman+15}. 
This also supports the scenario in which the cores are formed by gravitational contraction from moderate amplitudes rather than by strong shocks, since in the latter case a density discontinuity would be expected, implying a steep density gradient in all cases.

\subsection{The infall velocity profile and core ``coherence''} \label{sec:vinf}

It is also important to note that, according to eq.\ (\ref{eq:v_inflow}), $v_r \rightarrow 0$ as $r \rightarrow 0$ for $0<p<2$. That is, for slopes shallower than $-2$, the infall speed {\it decreases} towards the center. This is qualitatively (albeit not quantitatively) consistent with the prestellar similarity solution \citep{Larson69, Penston69, Whitworth+85}, for which the infall speed is linear with radius in the central, flat region. Instead, it is contrary to the inside-out (proto-stellar) solution of \citet{Shu77}, in which the infall speed {\it increases} towards the center as $r^{-1/2}$, and which applies only for the {\it protostellar} (post-singularity) stage of the collapse.

The inwards decrease of the infall speed during the prestellar stage implies that, as one samples the core at smaller scales and higher densities, the measured velocities will also be smaller. If this is reflected in the linewidth of the region, the measured nonthermal contribution to the linewidth will be smaller for more central regions, in agreement with observations \citep[e.g.,] [] {Goodman+98, Pineda+10, Chen+19, Chen+20}. This inwards decrease of the nonthermal contribution to the linewidth is referred to as ``coherence'', and interpreted in terms of turbulent dissipation. Instead, here we interpret it simply as a consequence of the inwards decrease of the infall speed in prestellar cores.

An additional point to note here is that, if the decrease of the nonthermal contribution to the linewidth were really due to the dissipation of turbulence, then one should be able to find a significant fraction of prestellar cores with sizeable velocity dispersions, corresponding to early stages in which the turbulence has not been dissipated yet.

Finally, we remark here that, the more advanced the collapse, the smaller the central Jeans-stable region, whose size is of the order of the Jeans length \citep{Keto+10} for the corresponding central density. This leads to the prediction that one should find an inverse correlation between the central density $\rho_{\rm c}$ and the size $R_{\rm c}$ of the
constant-density region of ``coherent'' cores of the form $R_{\rm c} \propto \rho_{\rm c}^{-1/2}$.

\subsection{Analogy with the mass-to-flux ratio in the magnetic case} \label{sec:magnetic}

A very important analogy of the mechanism discussed here (that the smallest scales are the last ones to appear unstable during collapse) occurs in the case of magnetic support. In this case, it is well known that a cloud or core can be supported by the magnetic field if it has a {\it subcritical} mass-to-magnetic flux ratio, where the critical value is $(M/\phi)_{\rm cr} = \alpha G^{-1/2}$, $G$ being the gravitational constant and $\alpha$ a geometrical constant \citep{Strittmatter66}. Hereinafter, we denote the mass-to-flux ratio of a cloud or core, normalized to the critical value, by $\mu$. 

For clouds of fixed mass, it is well known that the magnetic support is absolute, meaning that the value of $\mu$ remains constant as a cloud contracts or expands.
However, the situation is less obvious when fragments of a cloud or variable-mass clouds are considered. In the case of fragments (i.e., subregions) of a cloud, \citet{VS+05} showed that the mass-to-flux ratio of a fragment, $\mu_{\rm fr}$, must satisfy $\mu_{\rm fr} \le \mu_{\rm cl}$, where $\mu_{\rm cl}$ is the mass-to-flux ratio of the parent cloud. They showed this by considering two limiting cases. On one hand, they considered the limiting case of a uniform-density cloud of size $R_{\rm cl}$, and a subregion of size $R_{\rm fr}$ within it. In this case, the mass of the subregion scales as $R^3$, while the flux through it scales as $R^2$, and so the mass-to-flux ratio of the subregion is related to that of the cloud by $\mu_{\rm fr} = \mu_{\rm cl} R_{\rm fr}/R_{\rm cl}$. On the other hand, they considered the opposite limit in which the whole cloud contracts from size 
$R_{\rm cl}$ to size $R_{\rm fr}$. In this case, under ideal MHD, both the mass and the flux are conserved, and thus $\mu_{\rm fr} = \mu_{\rm cl}$. Any intermediate case, in which the fragment has a higher density than that of its parent cloud but a smaller mass, must have a mass-to-flux ratio intermediate between these limiting cases, and so it must satisfy
\begin{equation}
\mu_{\rm cl} \left(\frac{R_{\rm fr}} {R_{\rm cl}}\right) \le \mu_{\rm fr} \le \mu_{\rm cl}.
\label{eq:mu_fr_mu_cl}
\end{equation}
This shows that the mass-to-flux ratio {\it measured} for any fragment of a cloud should in general be smaller than that of its parent cloud, as long as ideal MHD holds. This result was verified numerically by \citet{Lunttila+09} and observationally by \citet{Crutcher+09}. It could also explain why $\mu$ is found to be $\sim1$ or even $<1$ in collapsing massive dense cores, such as those reported in \citet[][]{Palau+20}, or A\~nez-L\'opez et al. (2020)
\citep[see also][]{2020arXiv201213060A, 2020arXiv201204297B}.

We can now consider the case of our collapsing cores with time-varying power-law density profiles. In this case, we assume a density profile given by eq.\ (\ref{eq:dens_prof}),  which implies that the mean density within radius $r$ follows a scaling with the same exponent, as given by eq.\ (\ref{eq:mean_rho}). For convenience, we rewrite eq.\ (\ref{eq:mean_rho}) as $\rhoavg = \rhoavg_0 (r/r_0)^{-p}$, with $\rhoavg_0 \equiv 3 \rho_0/(3-p)$. We also assume that the mean magnetic field strength within the core scales with the mean density $\rhoavg$ of a core as 
\begin{equation}
B(\rhoavg) = B_0 \left( \frac{\rhoavg} {\rhoavg_0} \right)^\beta.
\end{equation}
Therefore, the dimensional mass-to-flux ratio out to radius $r$ is given by
\begin{equation}
    \frac{M}{\phi}(r) = \frac{4 \pi \rhoavg r^3/3}{\pi B_0 (\rhoavg/\rhoavg_0)^\beta r^2} = \frac{4 \rhoavg_0 r^{1-p(1-\beta)}} {B_0 r_0^{-p(1-\beta)}}.
    \label{eq:M2F_n}
\end{equation}

Of particular interest is the case $\beta = 2/3$, which is the scaling expected for a spherical mass contracting while conserving magnetic flux \citep[see, e.g.,] [] {Shu92}. In this case, we find that $M/\phi \propto R^{1-p/3}$, so that it decreases monotonically with decreasing radius for all physically plausible values of $p$. This implies that, similarly to the case of thermal support, magnetic support also appears stronger in the innermost parts of a cloud, { which may even appear magnetically subcritical (i.e., magnetically supported) even if they are just the central parts of a globally collapsing supercritical cloud.}

\subsection{The virial theorem perspective} \label{sec:VT}

{ Although fully dynamical and out-of-equilibrium, our results are consistent with the virial theorem (VT), albeit in a fashion different from how it is most frequently considered. 

\citet{McKee_Zweibel92} investigated the {\it Eulerian} form of the VT, which is the most relevant for our treatment with a fixed boundary at $r = r_0$, across which mass and energy can flow. This form reads
\begin{eqnarray}
\frac{1}{2} \frac{d^2I} {dt^2}  &=& \Big(2K - 2 \oint_S \xx \cdot \rho \uu \uu \cdot d\s - \frac{d} {dt} \oint_S \rho x^2 \uu  \cdot d\s \Big) \nonumber\\
&+& \Bigl(2U - \oint_S P \xx \cdot d\s \Bigr) \nonumber\\
&+& \Bigl(\M + \oint_S \xx \cdot \T \cdot d\s\Bigr) + W,
\label{eq:VT}
\end{eqnarray}
where $V$ is the volume of a region in the medium, $S$ is the surface enclosing this volume, $\xx$ is the position vector, $\uu$ is the velocity vector, $\bm B$ is the magnetic field, $I = \int_v \rho x^2 dV$ is the moment of inertia of the mass within volume $V$, $K = 1/2\, \int_V \rho u^2 dV$ is the nonthermal kinetic energy, $U = 3/2\, \int_V P\, dV$ is the internal energy, $\M = \int_V B^2/8\pi\, dV$ is the magnetic energy, $\T$ is the Maxwell stress tensor, and $W < 0$ is the gravitational energy. The terms within brackets on the right-hand side of the equation group sets of terms referring to specific physical agents (velocity field, thermal pressure, and magnetic field), for each of which there is a volume contribution and one or two surface contributions. The latter denote the work done on surface $S$ by the corresponding stresses. The last term within the first set of brackets can be interpreted as the rate of change of the moment of inertia flux across surface $S$ \citep[e.g.,] [] {BP+99}.

The VT is most often applied to {\it equilibrium} cases, for which $d^2I/dt^2 = 0$. Also, the surface terms are often neglected, which amounts to assuming that the gas within volume $V$ (often considered to be the cloud or clump) is isolated, with its surroundings producing negligible effects. In this case, the support against gravity provided by the internal or magnetic energies is derived by equating the volume terms corresponding to these energies to the gravitational energy ($2U = -W$ or $\M = -W$, respectively), leading to the standard Jeans and critical mass-to-magnetic flux criteria. Also, under the assumption that the nonthermal kinetic energy corresponds to random turbulence that provides support against gravity, equating $2K$ to $-W$ leads to a "turbulent Jeans criterion" \citep[e.g.,] [] {MacLow_Klessen04}.

However, for a dynamical collapse {\it flow} (rather than a fixed mass) considered within fixed boundaries as we do here, neither of the above assumptions is applicable, and thus the complete virial formulation of the problem must include both the time derivative and surface terms, modifying the standard criteria. Indeed, when an external, compressive velocity field is taken into account, the effective Jeans length (or mass) is reduced \citep{Hunter79, Hunter_Fleck82}, in agreement with our finding that scales smaller than the Jeans length are nevertheless contracting in the central parts of the collapse flow. This can be interpreted either as a reduction of the effective Jeans mass in the presence of an external compressive velocity field, or as the dynamic compression by ``ram pressure'' of an otherwise gravitationally stable core. Similarly, \citep{Guerrero_Vazquez20} have recently shown that, during the prestellar stage of gravitational collapse, a virial-like ratio of turbulent to infall energy can arise, albeit with no implication of turbulent support nor of a hydrostatic state. This can happen because of the dissipation of turbulent energy (the work done by the viscous forces), for which another term must also be included in the VT. Finally, during the prestellar collapse process, the moment of inertia of the cloud or core is in general changing in time as the object becomes more centrally condensed and its density profile steepens. In conclusion, our results are fully consistent with the VT when all the terms relevant for an evolutionary collapse process are considered.

}

\section{Summary and conclusions} \label{sec:concls}

In this paper we have calculated the approximate evolution of the radial density profile slope during the prestellar evolution of a core that starts as a moderate density perturbation on top of a uniform background. For simplicity, we have approximated the slope of the profile by a single power law of the form $\rho \propto r^{-p}$, where $p$ is allowed to vary. We recover the result by \citet{Li18} that the conditions that {\it a)} the infall speed at every radius is the gravitational velocity, eq.\ (\ref{eq:vr}) and {\it b)} the radial mass flux across spherical shells is independent of radius, require $p=2$. Furthermore, we have shown that this slope is an attractor because, as indicated by eq.\ (\ref{eq:dpdt}), values of $p<2$ cause $p$ to increase with time, while values of $p>2$ cause $p$ to decrease. Therefore, when the collapse starts from a very mild density enhancement, characterized by $p \gtrsim 0$, the evolution drives the slope towards $p=2$, in agreement with numerical simulations of this configuration \citep{NaranjoRomero+15}. This result is inaccessible to similarity studies, which apply to the case where the initial and boundary conditions are infinitely far in the temporal and spatial domains from the range of interest in the system, and therefore cannot address the evolution during the early transient stages. 

We also discussed the fact that, within any prestellar contracting core, there always exists an inner region that is smaller than the Jeans length, and so it is, in effect, not self-gravitating. However, it constitutes the ``tip of the iceberg'' of the entire collapsing structure, and is being compressed (``crushed'') by the ram pressure of the infalling outer envelope of the core, in which the power-law regime applies, and which is at lower density than the inner, Jeans-stable region. This implies that the observation of Jeans stability of the central dense cores is not an indication that they are hydrostatic and, as hinted by \citet{Kirk+17}, the observed external pressure is likely to actually be compressive ram pressure rather than ``confining'' thermal or microturbulent pressure.

We then considered the infall velocity profile, showing that it tends to zero at the core center for $p<2$, thus offering an explanation for the observed ``velocity coherence'' of moderate-density prestellar cores, already pointed out in \citet{NaranjoRomero+15}: in this view, the decrease of the nonthermal velocity dispersion in the innermost regions of the cores is just a consequence of the inwards decrease of the infall speed, rather than the dissipation of any supporting microturbulence. In fact, if turbulent dissipation were the reason for coherence, one would expect to see a distribution of velocity dispersions for starless cores of a given central density, with a significant fraction of them exhibiting supersonic velocity dispersions, corresponding to cores observed before they managed to dissipate their supporting turbulence. This is not observed in the studies of \citet{Chen+19} and \citet{Li+20}, where most of the starless cores have subsonic or transonic velocity dispersions. Therefore, the interpretation in terms of the decreasing inwards velocity appears preferable over that in terms of the dissipation of turbulence.

Another implication of our results is that the infall velocity extends beyond the point where the core meets the uniform background, implying accretion {\it onto the core}, which is not considered in the standard inside-out collapse model \citep{Shu77}. In the latter, the dynamical collapse starts at the time of the formation of the singularity, at the tip of the SIS. After that, the envelope remains at rest, and only the inner region undergoes collapse. Instead, if onset of the collapse occurs when the density fluctuation is only moderate (i.e., significantly earlier than the time of formation of the singularity, contrary to the assumption in the inside-out collapse model), the rarefaction front has been expanding for one large-scale free-fall time, and thus the infall motions extend much further out than in Shu's mode, in agreement with observations \citep{Lee+01}, and implying accretion onto the core, in agreement with observations that accretion onto protostar is fed from the larger clump scale \citep[e.g.] [] {Liu+15, Ohashi+16, Yuan+18, Peretto+20}.

We also briefly examined the case of magnetic support, and showed that a similar phenomenon appears: the mass-to-magnetic flux decreases towards the innermost parts of a magnetically-supported structure under ideal MHD conditions, and thus the centermost parts may appear magnetically supported (subcritical) even when they may be part of a magnetically supercritical large-scale structure undergoing global gravitational contraction.

Finally, we briefly recalled the available observational evidence supporting our result that the density profile steepens during the growth of the core by gravitational infall, so that, when the central structure still appears far from being locally gravitationally bound, the density profile is rather shallow \citep[$p\sim 1$] [] {Chen+19}, while in objects that already contain a YSO, the slope \citep[$p \sim 2$] [] {Palau+14}. 

Our results, together with the available observational evidence, support the scenario of global hierarchical collapse (GHC) in molecular clouds, in which a continuous gravitationally driven accretion flow occurs in the clouds, consisting of a hierarchy of collapses within collapses, each scale accreting from the next larger one, and smaller-scale collapses starting later, but finishing earlier than the cloud scale one \citep{VS+19}.

\section*{Acknowledgements}

{ The authors wish to thank Pierre Didelon and the reviewer of this manuscript for their useful comments.}
This project has received financial support from CONACYT grant 255295 to E.V.-S. 
A.P. acknowledges financial support from CONACyT and UNAM-PAPIIT IN113119 grant, M\'exico.

\section*{Data availability}

The data underlying this article will be shared on reasonable request to the corresponding author.




\bibliographystyle{mnras}
\bibliography{collapse}



\bsp	
\label{lastpage}
\end{document}